\documentstyle[aps,axodraw,psfig,eqsecnum]{revtex}

\title{\bf Chiral Gauge Theory on Lattice with Domain Wall Fermions}
\author{{S. Basak\thanks{Permanent Address: Dept. of Physics, NND College,
Calcutta 700 092}} and {Asit K. De}}
\address{Theory Group, Saha Institute of Nuclear Physics, 1/AF Salt 
Lake, Calcutta 700064, India}
\date{\today}

\begin{document}
\maketitle

\begin{abstract}
We investigate a $U(1)$ lattice chiral gauge theory ($L\chi GT$) with domain 
wall fermions and compact
gauge fixing. In the {\em reduced} model limit, our perturbative and numerical
investigations show that there exist no extra mirror chiral modes. The 
longitudinal gauge degrees of freedom have no effect on the free 
domain wall fermion spectrum consisting of opposite chiral modes at the 
domain wall and at the anti-domain wall which have an exponentially 
damped overlap. 
\end{abstract}

\section{Introduction} \label{intro}
Lattice regularization of chiral gauge theories has remained a long standing
problem of nonperturbative investigation of quantum field theory. Lack of
chiral gauge invariance in $L\chi GT$ proposals is responsible for
the longitudinal gauge degrees of freedom ({\em dof}) coupling to
fermionic {\em dof} and eventually spoiling the chiral nature of the theory.
The well-known example is the Smit-Swift proposal of $L\chi GT$ \cite{smit}.
Although in a recent development using a Dirac operator that
satisfies the Ginsparg-Wilson relation, it was possible to formulate a
$L\chi GT$ without violating gauge-invariance or locality \cite{luscher}, an
explicit model for nonperturbative numerical studies is still not 
available.
 
In this paper, we follow the gauge fixing approach to $L\chi GT$
\cite{golter1}. The obvious remedy to control the longitudinal gauge {\em dof}
is to gauge fix with a target theory in mind. The Roma proposal \cite{roma}
involving gauge fixing passed perturbative tests but does not address the
problem of gauge fixing of compact gauge fields and the associated problem of
lattice artifact Gribov copies. The formal problem is that
for compact
gauge fixing a BRST-invariant partition function as well as (unnormalized)
expectation values of BRST invariant operators vanish as a consequence of
lattice Gribov copies \cite{neuberger}. Shamir and Golterman \cite{golter1}
have proposed to
keep the gauge fixing part of the action BRST noninvariant and tune
counterterms to recover BRST in the continuum. In their formalism, the
continuum limit is to be taken from within the broken ferromagnetic (FM) phase
approaching another broken phase which is called ferromagnetic directional
(FMD) phase, with the mass of the gauge field vanishing at the FM-FMD
transition. This was tried out in a $U(1)$ Smit-Swift model and so
far the results show that in the pure gauge sector, QED is recovered in
the continuum limit \cite{recent} and in the {\em reduced} model limit 
(to be defined below) free chiral fermions in the appropriate chiral 
representation are obtained \cite{bock1}. Tuning with counterterms has 
also not posed any practical problem, actually very little tuning is 
necessary. Efforts are currently underway to extend this gauge fixing 
proposal to include nonabelian gauge groups \cite{nab}. 
      
Without gauge fixing the longitudinal gauge {\em dof}, which are 
radially frozen scalar fields, are rough and nonperturbative even if the 
transverse gauge coupling may be weak (this is because with the standard 
lattice measure, each point on the gauge orbit has equal weight). 
The theory in the continuum limit, taken at the transition 
between the broken symmetry ferromagnetic (FM) phase and the 
symmetric paramagnetic (PM) phase, displays undesired nonperturbative 
effects of the scalar-fermion coupling that usually spells disaster for 
the chiral theory. The job of the gauge fixing is to introduce a new 
continuous phase transition, from the FM phase to a new broken symmetry 
phase (FMD), at which the gauge symmetry is recovered and at the same 
time the gauge fields become smooth.
  
The problem can be cleanly studied in the reduced model as explained in the
following. When one gauge transforms a gauge non-invariant theory, one picks
up the longitudinal gauge degrees of freedom (radially frozen scalars)
explicitly in the action. The reduced model is then obtained   
by making the lattice gauge field unity for 
all links, {\em i.e.}, by switching off the transverse gauge coupling. 
The action becomes that of a chiral Yukawa theory with interaction 
between the fermions and the longitudinal gauge {\em dof}. 
The reduced model would have a phase structure similar to the full 
theory, {\em e.g.}, the gauge fixed theory in the reduced limit 
will have a FM-FMD transition in addition to the FM-PM transition. 
Now for the gauge fixing proposal to work, the scalars need 
to decouple from the fermions at the FM-FMD transition leaving the 
fermions free in the appropriate chiral representation. Passing the 
reduced model test is an important first step for any $L\chi GT$ 
proposal that breaks gauge invariance.  

In the reduced model derived from the gauge fixed theory the 
scalar fields become smooth and expandable in a perturbative 
series as $1+{\cal O}(\mbox{coupling constant})$ at the FM-FMD 
transition. If continuum limit can be taken near the point in the 
coupling parameter space around which this perturbative expansion is 
defined, the scalar fields will decouple from the theory. 
The parameterization of the gauge fixing action turns out to be a good 
one, because this continuum limit can be taken ($i$) very easily by 
approaching the FM-FMD transition almost perpendicularly by tuning 
essentially one counterterm, and ($ii$) at a point on this transition 
line which is reasonably far away from the expansion point. This has 
been possible in \cite{bock1} and again in the present work.  

A central claim of the gauge fixing proposal 
is that it is universal, {\em i.e.}, it should work with any lattice 
fermion action that has the correct classical continuum limit. This is 
because the central idea as discussed above is independent of the 
particular lattice fermion regularization. In the 
present paper we want to confirm the universality claim by applying the 
proposal to domain wall fermions \cite{kaplan} with $U(1)$ gauge group. 
For this purpose we have chosen the waveguide formulation \cite{kaplan2} 
of the domain wall fermion and investigate in the reduced model. This 
model was investigated before without gauge fixing and the free domain 
wall spectrum was not obtained in the reduced limit \cite{golter2}.  
Mirror chiral modes were found at the waveguide boundaries in addition 
to the chiral modes at the domain wall or anti-domain wall.
 
In section II we present the gauge-fixed domain wall fermion action for a
$U(1)$ chiral gauge theory and then go to the so-called reduced
model by switching off the transverse gauge coupling. In section III we
perform a weak coupling perturbation theory in the reduced model for the
fermion propagators and mass matrix to 1-loop. However, in sections
\ref{ferm_m} and \ref{ferm_m1} we have used special boundary conditions
(instead of the actual Kaplan boundary conditions) to arrive at explicit
expressions for the overlap of the opposite chiral modes. Our numerical 
results for 
the quenched phase diagram and chiral fermion propagators at the domain 
wall and anti-domain wall and at the waveguide boundaries are presented 
and compared with the perturbative results in section IV. We summarize 
in the concluding section V. In Appendix A, we describe the special 
boundary conditions used in sections \ref{ferm_m} and \ref{ferm_m1}.
In Appendix B we schematically discuss how using Kaplan boundary conditions
one can arrive at the same qualitative conclusion about the 1-loop
overlap of the opposite chiral modes.
 
\section{Gauge-fixed Domain Wall Action} \label{gfdwa}

Kaplan's free domain wall fermion action \cite{kaplan} on a
$4+1$-dimensional lattice is given by (lattice constant is taken to be
unity throughout this paper),
\begin{equation}
S_F = \sum_{XY} \overline{\psi}_X \left[ \partial\!\!\!/_5 - w_5 +
{\bf M}\right]_{XY} \psi_Y  \label{dwact}
\end{equation}
where $\overline{\psi}$ and $\psi$ are the fermion fields, and  
$\partial\!\!\!/_5$ and $w_5$ are respectively the 5-dimensional
Dirac operator and the Wilson term,
\begin{eqnarray}
(\partial\!\!\!/_5)_{XY} &=&
\frac{1}{2}\sum_{\alpha=1}^5 \gamma_\alpha \left( 
\delta_{X+\hat{\alpha},Y} - \delta_{X-\hat{\alpha},Y} \right),            
\nonumber \\ (w_5)_{XY} &=& \frac{r}{2} \sum_{\alpha=1}^5
\left( \delta_{X+\hat{\alpha},Y} + \delta_{X-\hat{\alpha},Y} - 2\delta_{XY}
\right),  \label{par5}
\end{eqnarray}
The $\gamma_\alpha$'s are the five hermitian euclidean gamma matrices,
$r$ is the Wilson parameter, $X=(x,s),~Y=(y,t)$ label the sites of
the $L^4 L_s$ lattice and $L_s$ is the extent of the 5th dimension:
$0 \leq s,t \leq L_s-1$. We are interested in
taking the continuum limit in the 4 space-time dimensions only. It is
convenient to look at the 5-th dimension as a flavor space.

With periodic boundary conditions in the 5th or
$s$-direction  ($s,t = L_s \Rightarrow s,t=0$) and the domain wall mass
${\bf M}$ taken as
\begin{equation}
{\bf M}_{XY}=m(s)\delta_{XY},\;\; {\rm where},
\end{equation}
\begin{equation}
m(s) =  \begin{array}{rl}
-m_0, &  0 <s< L_s/2 \\
 0,   &  s = 0, L_s/2 \\
 m_0, &  L_s/2 <s< L_s
\end{array} \label{dwmass}
\end{equation}
the model possesses a lefthanded (LH) chiral mode bound to the domain wall
at $s=0$ and a righthanded (RH) chiral mode bound to the anti-domain wall
at $s=L_s/2$. For $m_0 L_s\gg 1$, these modes have exponentially
small overlap. The chiral modes exist for momenta $p$
below a critical momentum $p_c$, i.e. $\vert \hat{p} \vert < p_c$, where
$\hat{p}^2 = 2\sum_\mu[1-\cos(p_\mu)]$ and $p_c^2 = 4 - 2m_0/r$. Taking
the Wilson parameter $r=1$ the choice of $m_0$ is then restricted to
$0<m_0<2$.

A 4-dimensional gauge field which is same for all
$s$-slices can be coupled to fermions only for a restricted number of
$s$-slices around the anti-domain wall \cite{golter2} with a view to
coupling only to the RH mode at the anti-domain wall. The gauge field
is thus confined within a {\em waveguide},
\begin{eqnarray}
WG &=& (s: s_0 < s \leq s_1) \nonumber \\
{\rm with} \hspace{0.5cm} s_0 &=& \frac{L_s+2}{4}-1, \hspace{0.7cm} s_1 =
\frac{3L_s+2}{4}-1.
\end{eqnarray}
With this choice, $(L_s-2)$ has to be a multiple of 4. For convenience,
the boundaries at ($s_0,s_0+1$) and ($s_1,s_1+1$) are denoted waveguide
boundary $I$ and $II$ respectively.

The gauge transformations on the fermion fields are defined as follows:
\begin{eqnarray}
\psi_x^s \rightarrow g_x\psi_x^s, \hspace{0.8cm} &\overline{\psi}_x^s &
\rightarrow \overline{\psi}_x^s g_x^\dagger \hspace{0.8cm} s \in WG
\nonumber \\
\psi_x^s \rightarrow \psi_x^s, \hspace{0.8cm} &\overline{\psi}_x^s&
\rightarrow \overline{\psi}_x^s \hspace{1.15cm} s \in\!\!\!\!\!/ \;WG
\label{gsym}
\end{eqnarray}   
where $g_x\in G$, the gauge group. Other symmetries of the model remain 
the same as in \cite{golter2}.

Obviously, the hopping terms from $s_0$ to $s_0+1$ and that
from $s_1$ to $s_1+1$ would break the local gauge invariance of the
action. This is taken care of by gauge transforming the action and
thereby picking up the pure gauge {\em dof} or a radially frozen scalar
field $\varphi$ (St\"{u}ckelberg field) at the waveguide boundary,
leading to the gauge-invariant action (with $\varphi_x\rightarrow g_x
\varphi_x$ and (\ref{gsym})) :
\begin{eqnarray}
S_{F} & = & \sum_{s \in WG}
\overline{\psi}^s \left( D\!\!\!\!/\,(U) - W(U) + m(s) \right) \psi^s
+ \sum_{s\not\in WG} \overline{\psi}^s \left( \partial\!\!\!/
- w + m(s) \right) \psi^s \nonumber \\
&+& \;\sum_s \overline{\psi}^s \psi^s - \sum_{s\neq s_0,s_1}
\left( \overline{\psi}^s P_L \psi^{s+1} + \overline{\psi}^{s+1} P_R
\psi^s \right) \nonumber \\
&-& \;y \left( \overline{\psi}^{s_0} \varphi^\dagger P_L \psi^{s_0+1} +
\overline{\psi}^{s_0+1} \varphi P_R \psi^{s_0} \right)
- y \left( \overline{\psi}^{s_1} \varphi P_L
\psi^{s_1+1} + \overline{\psi}^{s_1+1} \varphi^\dagger P_R \psi^{s_1}
\right) \label{wgact}
\end{eqnarray}
where we have taken the Wilson parameter $r=1$
and have suppressed all indices other than $s$. The projector $P_{L(R)}$
is $(1\mp \gamma_5)/2$ and $y$ is the Yukawa coupling introduced by hand at
the waveguide boundaries. $D\!\!\!\!/\;(U)$ and $W(U)$ are respectively the
gauge covariant Dirac operator and the Wilson term in 4 space-time dimensions.
$\partial\!\!\!/$ and $w$ are the 4-dimensional versions of (\ref{par5}).

The gauge-fixed pure gauge action for $U(1)$, where the ghosts are free
and decoupled, is:
\begin{equation}
S_B(U) = S_g(U) + S_{gf}(U) + S_{ct}(U) \label{ggact}
\end{equation}
where, $S_g$ is the usual Wilson plaquette action;
the gauge fixing term $S_{gf}$ (as proposed by Shamir and Golterman) and 
the gauge field mass counter term $S_{ct}$ are given by (for a 
discussion of relevant counterterms see \cite{golter1,bock2}),
\begin{eqnarray}
S_{gf}(U) & = & \tilde{\kappa} \left( \sum_{xyz} \Box(U)_{xy}
\Box (U)_{yz} - \sum_x B_x^2 \right), \label{gfact} \\
S_{ct}(U)  & = &  - \kappa \sum_{x\mu} \left( U_{\mu x} +
U_{\mu x}^\dagger \right),
\end{eqnarray}
where $\Box(U)$ is the covariant lattice laplacian and
\begin{equation}
B_x  = \sum_\mu \left( \frac{V_{\mu x-\hat{\mu}} +
V_{\mu x}}{2} \right)^2  \label{bxsq}
\end{equation}
with
$V_{\mu x} = \frac{1}{2i} \left( U_{\mu x} - U_{\mu x}^\dagger \right)$
and $\tilde{\kappa} = 1/(2\xi g^2)$.

$S_{gf}$ is not just a naive lattice transcription of the continuum 
covariant gauge fixing term, it has in addition appropriate irrelevant 
terms. As a result, $S_{gf}$ has a unique absolute minimum at $U_{\mu 
x}=1$, validating weak coupling perturbation theory (WCPT) around $g=0$ 
or $\tilde{\kappa}=\infty$ and in the naive continuum limit it reduces 
to $\frac{1}{2\xi} \int d^4x (\partial_\mu A_\mu)^2$. 

Obviously, the action $S_B(U)$ is not gauge invariant. By giving it a gauge
transformation the resulting action $S_B(\varphi^\dagger_x U_{\mu x}
\varphi_{x+\hat{\mu}})$ is gauge-invariant with $U_{\mu x}\rightarrow g_x U_{\mu
x} g^\dagger_{x+\hat{\mu}}$ and $\varphi_x \rightarrow g_x \varphi_x$, $g_x \in
U(1)$. By restricting to the trivial orbit, we arrive at the so-called
{\bf reduced model} action
\begin{equation}
S_{reduced} = S_F(U=1) + S_B(\varphi^\dagger_x \;1 \;\;\varphi_{x+\hat{\mu}})
\label{reduced}
\end{equation}
where $S_F(U=1)$ is obtained quite easily from eq.(\ref{wgact}) and
\begin{equation}
S_B(\varphi^\dagger_x \;1 \;\;\varphi_{x+\hat{\mu}})
= -\kappa \sum_x \varphi^\dagger_x(\Box \varphi)_x + \tilde{\kappa}
\sum_x \left[\varphi^\dagger_x(\Box^2 \varphi)_x - B^2_x \right] \label{redB}
\end{equation}
now is a higher-derivative scalar field theory action. $B_x$ in (\ref{redB})
is same as in (\ref{bxsq}) with
\begin{equation}
V_{\mu x} = \frac{1}{2i}\left(\varphi^\dagger_x\varphi_{x+\hat{\mu}} -
\varphi^\dagger_{x+\hat{\mu}} \varphi_x\right).
\end{equation}

In the following, we investigate the action (\ref{reduced}) at $y=1$ by
analytical and numerical methods. Some numerical results with other values of
$y$ have been presented in \cite{npbps}. The waveguide model strictly
at $y=0$ would give rise to opposite
chiral modes at the waveguide boundaries as can be seen from fermion current
considerations \cite{golter2} (and also from numerical simulation) 
and would thereby spoil the chiral nature of the theory. It is an 
interesting question to investigate the model for $0<y<1$. Analysis of 
the results for small values of $y (<1)$ is tricky and will be discussed 
in a separate article \cite{big}.

\section{Weak Coupling Perturbation Theory in the Reduced Model} \label{wcpt}
At $y=1$, we
carry out a WCPT in the coupling $1/\tilde{\kappa}$ for the fermion
propagators to 1-loop. In order to develop perturbation theory, in reduced
model, we expand,
\begin{equation}
\varphi_x = \exp(ib\theta_x) = 1 +ib \theta_x  -\frac{1}{2} b^2 \theta_x^2
+ {\cal O}(b^3)
\end{equation}
where,
$b=1/\sqrt{2\tilde{\kappa}} $  and $\theta_x$ is dimensionless, 
leading to
\begin{equation}
S = S_F^{(0)}(\psi,\overline{\psi};y) + S_B^{(0)}(\theta) + S^{({\rm
int})}(\psi,\overline{\psi},\theta;y)  \label{Sint}
\end{equation}
where $S^{(0)}$'s are free actions and $S^{({\rm int})}$ is the interaction
part.

\subsection{Scalar propagator at tree level} \label{scal_p}
From $S_B^{(0)}(\theta)$ one gets the free propagator for the compact scalar
$\theta$ \cite{bock2},
\begin{equation}
{\cal G}(k) = \frac{1}{\hat{k^2}(\hat{k^2} + \omega^2)},
\hspace{1.0cm} \omega^2=\frac{\kappa}{\tilde{\kappa}}
\end{equation}
where, $\hat{k}_\mu = 2\sin(k_\mu/2)$.

\subsection{$LL$ and $RR$ fermion propagators at tree level and at 1-loop}
\label{ferm_p}
\subsubsection{Tree level}
With $y=1$, $S_F^{(0)}(\psi,\overline{\psi};y=1)$ is the free domain wall
action (\ref{dwact}).
Free fermion propagators at $y=1$
are obtained in momentum space for 4-spacetime dimensions while staying
in the coordinate space for the 5th dimension following \cite{aoki1}
(results in \cite{aoki1,aoki2,blum} cannot be directly used because of
difference in implementation of the domain wall (\ref{dwmass})). The
free action is written as,
\begin{equation}
S_F^{(0)}(y=1) = \sum_{p,s,t} \overline{\widetilde{\psi_p^s}} \left[
i\overline{p\!\!\!/} \delta_{s,t} + M_{st}P_L + M^\dagger_{st} P_R
\right] \widetilde{\psi_p^t}
\end{equation}
where, $M_{st} = F(p)\delta_{s,t} + (M_0)_{st}$, $(M_0)_{st} =
[1+m(s)]\delta_{s,t} - \delta_{s+1,t}$,
$F(p) = \sum_\mu (1 - \cos(p_\mu))$,
$\overline{p}_\mu = \sin(p_\mu)$ and $\overline{p\!\!\!/} =
\gamma_\mu\overline{p}_\mu$.
The free fermion propagator can formally be written as,
\begin{eqnarray}
\Delta(p) &=& \left[ i\overline{p\!\!\!/} + M P_L +
                                  M^\dagger P_R \right]^{-1} \nonumber \\
 &=& ( -i\overline{p\!\!\!/} + M^\dagger ) P_L G_L(p) +
   ( -i\overline{p\!\!\!/} + M ) P_R G_R(p)  \label{fprop}
\end{eqnarray}
where,
\begin{eqnarray}
G_L(p) & = & \frac{1}{\sum_\mu \overline{p}_\mu^2 + M M^\dagger}
\label{gl} \\
G_R(p) & = & \frac{1}{\sum_\mu \overline{p}_\mu^2 +
M^\dagger M}. \label{gr}
\end{eqnarray}
Solution of $G_L$ is obtained by writing (\ref{gl}) explicitly:
\begin{equation}
\left[ \overline{p}^2+1+B(s)^2 \right] (G_L)_{s,t} - B(s+1)
(G_L)_{s+1,t} - B(s) (G_L)_{s-1,t} = \delta_{s,t} \label{green}
\end{equation}
and similarly for $G_R$. In (\ref{green}), $B(s)= F(p)+1+m(s)$.
We show only the calculations for obtaining $G_L$ and henceforth drop
the subscript $L$.

Setting the notation as follows:
\begin{eqnarray}
G &=& G^-, \;\;\;\;B(s)=F(p)+1-m_0=a_- \hspace{0.8cm} {\rm for}\;\;\;\;
0 <s \leq L_s/2-1 \label{nomen1}\\
{\rm and} \hspace{0.4cm} G &=& G^+, \;\;\;\;B(s)=F(p)+1+m_0=a_+
\hspace{0.8cm} {\rm for}\;\;\;\;L_s/2 <s\leq L_s-1, \label{nomen2}
\end{eqnarray}
the equations for $G^\pm$ are given by,
\begin{eqnarray}
\left( \overline{p}^2+1+a^2_{-} \right) G^{-}_{s,t} - a_{-} G^{-}_{s+1,t}
- a_{-} G^{-}_{s-1,t} &=& \delta_{s,t}, \label{lmslice} \\
\left( \overline{p}^2+1+a^2_{+} \right) G^{+}_{s,t} - a_{+} G^{+}_{s+1,t}
- a_{+} G^{+}_{s-1,t} &=& \delta_{s,t}. \label{lpslice}
\end{eqnarray}

The ranges of $s$ in eqs.(\ref{nomen1}, \ref{nomen2}) for which $G^-$
and $G^+$ are defined, are applicable
only to the translationally invariant eqs.(\ref{lmslice},
\ref{lpslice}). In general for the translationally noninvariant
eq.(\ref{green}) we also define $G^-$ and $G^+$ at $s=0$ and $L_s/2$,
the ones excluded by (\ref{lmslice}, \ref{lpslice}). The $\pm$
superscript to $G$ at $s=0,\;L_s/2$ is decided by the translationally
invariant $s$-sector from which $s=0$ or $L_s/2$ is approached in
eq.(\ref{green}). We have used this notation for the boundary conditions
eq.(\ref{dwbc}) below.

The solutions of the eqs.(\ref{lmslice}, \ref{lpslice}) are expressed
as sum of homogeneous and inhomogeneous solutions:
\begin{equation}
G^{\pm}_{s,t}(p) = g_{\pm}^{(1)}(t)e^{-\alpha_{\pm}(p)s} +
                   g_{\pm}^{(2)}(t)e^{\alpha_{\pm}(p)s}
+ \frac{\cosh[\alpha_{\pm}(p)(\vert s-t \vert - l/2)]}
{2a_{\pm}\sinh(\alpha_{\pm}(p))\sinh(\alpha_{\pm}(p)l/2)}, \label{lmsol}
\end{equation}
where, $l=L_s/2$ and
\begin{equation}
\cosh(\alpha_{\pm}(p))=\frac{1}{2}\left(
a_{\pm}+\frac{1+\overline{p}^2} {a_{\pm}} \right).
\end{equation}
The third term in (\ref{lmsol})
is the inhomogeneous solution. To avoid singularities in
$\alpha_{\pm}(p)$ when $a_{\pm}$ is zero further restricts the allowed range
of $m_0$ to $0<m_0<1$. In this paper we have taken $m_0=0.5$.

In order to get the complete
solution we need to determine the unknown functions $g_{\pm}^{(1)}(t)$
and $g_{\pm}^{(2)}(t)$ in (\ref{lmsol}), which are obtained by
considering boundary conditions from eqs.(\ref{lmslice}, \ref{lpslice})
at $s=0,\;1,\;L_s/2-1,\;L_s/2,\;L_s/2+1,\;L_s-1$,
\begin{eqnarray}
a_0 G^{-}_{0,t}(p) &=& a_{+} G^{+}_{L_s,t}(p), \nonumber \\
a_0 G^{+}_{L_s/2,t}(p) &=& a_{-} G^{-}_{L_s/2,t}(p), \nonumber \\
(\overline{p}^2+1+a_0^2) G^{-}_{0,t}(p) - a_{-}G^{-}_{1,t}(p)
&=& \delta_{0,t} + a_0 G^{+}_{L_s-1,t}(p), \nonumber \\
(\overline{p}^2+1+a_0^2) G^{+}_{L_s/2,t}(p) - a_{+}G^{+}_{L_s/2+1,t}
&=& \delta_{L_s/2,t} + a_0 G^{-}_{L_s/2-1,t}. \label{dwbc}
\end{eqnarray}
with $B(s)=F(p)+1=a_0$ at $s=0,L_s/2$. It is to be
noted that these boundary conditions are significantly different from
the ones given in \cite{aoki1} because of the difference in implementation
of the domain wall. $G^-_{0,t}$ and $G^+_{L_s/2,t}$ (and the corresponding
ones from $G_R$) are used to determine
the free chiral propagators at the domain wall and anti-domain wall for
comparison with numerical data in Fig.\ref{waw}. We will see later that
these chiral propagators do not receive any 1-loop self-energy corrections.

Substituting
$\overline{p}^2+1+a_0^2=F_0$ and $2a_{\pm}\sinh(\alpha_{\pm}(p))
\sinh(\alpha_{\pm}(p)l/2)=X_{\pm}$ and using the boundary conditions,
eqs.(\ref{dwbc}), we arrive at an equation of the form,
\begin{equation}
{\bf A}\cdot{\bf g}(t) = {\bf X}(t), \label{mateq}
\end{equation}
where, ${\bf g}(t) = (g_{-}^{(1)}\;\;\;g_{-}^{(2)}\;\;\;g_{+}^{(1)}\;\;\;
g_{+}^{(2)})$ is a 4-component vector, ${\bf X}(t)$ is another 4-component
vector and ${\bf A}$ is a $4\times 4$ matrix as given below,

\[ {\bf A} = \left( \begin{array}{cccc}
a_0 & a_0 & -a_{+}e^{-\alpha_{+}L_s} & -a_{+}e^{\alpha_{+}L_s} \\
a_{-}e^{-\alpha_{-}L_s/2} & a_{-}e^{\alpha_{-}L_s/2} & -a_0e^{-\alpha_{+}
L_s/2} & -a_0e^{\alpha_{+}L_s/2} \\
F_0-a_{-}e^{-\alpha_{-}} & F_0-a_{-}e^{\alpha_{-}} & -a_0e^{-\alpha_{+}
(L_s-1)} & -a_0e^{\alpha_{+}(L_s-1)} \\
-a_0e^{-\alpha_{-}(L_s/2-1)} & -a_0e^{\alpha_{-}(L_s/2-1)} &
(F_0-a_{+}e^{-\alpha_{+}})e^{-\alpha_{+}L_s/2} &
(F_0-a_{+}e^{\alpha_{+}})e^{\alpha_{+}L_s/2}
\end{array} \right) \]
and
\[ {\bf X}(t) = \left( \begin{array}{ll}
a_{+}X_{+}\cosh[\alpha_{+}(\vert L_s-t \vert -l/2)] & \hspace{-0.9cm} -\,
a_0X_{-}\cosh[\alpha_{-}(\vert -t \vert -l/2)] \\
& \\
a_0X_{+}\cosh[\alpha_{+}(\vert L_s/2-t \vert -l/2)] & \hspace{-0.7cm} -\,
a_{-}X_{-}\cosh[\alpha_{-}(\vert L_s/2-t \vert -l/2)] \\
& \\
\delta_{0,t} + a_0X_{+}\cosh[\alpha_{+}(\vert L_s-1-t \vert\, - &
\hspace{-0.3cm} l/2)] - F_0X_{-}\cosh[\alpha_{-}(\vert -t \vert -l/2)] + \\
& \hspace{3.5cm} a_{-}X_{-}\cosh[\alpha_{-}(\vert 1-t \vert -l/2)] \\
& \\
\delta_{L_s/2,t} + a_{+}X_{+}\cosh[\alpha_{+}(\vert L_s/2+1-&\!\!t \vert
-l/2)] - F_0X_{+}\cosh[\alpha_{+}(\vert L_s/2-t \vert -l/2)] + \\
& \hspace{2.4cm} a_0X_{-}\cosh[\alpha_{-}(\vert L_s/2-1-t \vert -l/2)]
\end{array} \right). \]

The explicit expressions for ${\bf A}$ and ${\bf X}(t)$ are obviously
different from similar expressions given in \cite{aoki1,aoki2,blum} because
of the differences in domain wall implementation as already discussed
earlier.

The solution to the eqs.(\ref{mateq}) is very complicated in general,
particularly for finite $L_s$. However, $g_{\pm}(t)$ can be obtained for
finite $L_s$ by solving the above equations numerically for different $t$
values. This way we can easily construct the free fermion propagators at any
given $s$-slice, including the zero mode propagators at $s=t=0,L_s/2$.

The solutions for $(G_R)_{s,t}$ and the resulting propagators are 
obtained in exactly the same way. However, in this case the explicit 
forms for (\ref{green}), (\ref{lmsol}), (\ref{dwbc}) and matrices ${\bf 
A}$ and ${\bf X}(t)$ in (\ref{mateq}) are obviously different.

\subsubsection{1-loop} \label{ferm_p1}
Next we calculate the chiral fermion propagators to 1-loop. {\em Half-circle}
diagrams which are diagonal in flavor space contributes to $LL$ and $RR$
propagator self-energies. However, the
self-energies are nonzero only at the waveguide boundaries $I$ and $II$.

Retaining up to ${\cal O}(b^2)$ in the interaction term
$S^{({\rm int})}(\psi,\overline{\psi},\theta;y=1)$ in (\ref{Sint}),
we find the vertices necessary to calculate the self-energies to
1-loop.

\vspace{-0.8cm}
\begin{center}
\begin{figure}
\begin{picture}(300,85)(0,65)
\ArrowLine(75,85)(125,85)
\ArrowLine(125,85)(175,85)
\ArrowLine(175,85)(225,85)
\DashArrowArcn(150,85)(35,180,0){3}
\Text(80,92)[]{\footnotesize{$L$}}
\Text(92,76)[]{$p$, $s_0+1$}
\Text(125,92)[]{\footnotesize{$R$}}
\Text(175,92)[]{\footnotesize{$R$}}
\Text(150,76)[]{$p-k$, $s_0$}
\Text(220,92)[]{\footnotesize{$L$}}
\Text(208,76)[]{$p$, $s_0+1$}
\Text(150,110)[]{$k$}
\end{picture}
\caption{1-loop self-energy contribution to $LL$ propagator at $WG$ boundary
$I$.} \label{hcirc}
\end{figure}
\end{center}

The $LL$ propagator on the $(s_0+1)$-th slice at the waveguide boundary
$I$ receives a nonzero  self-energy contribution from the
half-circle diagram,
\begin{eqnarray}
-\left(\Sigma_{LL}^I(p)\right)_{st} &= &\int_{BZ} \frac{d^4k}{(2\pi)^4}
b^2 \left[-i\gamma_\mu (\overline{p-k})_\mu P_L
G_L(p-k)\right]_{s_0,s_0} {\cal G}(k) \;\delta_{s,s_0+1}\delta_{t,s_0+1}
\label{fselfl} \\
& \rightarrow & \frac{b^2}{L^4} \sum_k \left[{\cal
S}^{(0)}_{RR}(p-k)\right]_{s_0,s_0}\;\; \frac{1}{\hat{k}^2 
(\hat{k}^2+\omega^2)} \;\delta_{s,s_0+1}\delta_{t,s_0+1} \label{fselfd}
\end{eqnarray}
where  the expression in the square bracket in
(\ref{fselfl}) is the free $RR$ propagator 
$[{\cal S}^{(0)}_{RR}]_{s_0,s_0}$ on the $s_0$-slice. 
Eq.(\ref{fselfl}) assumes 
infinite 4 space-time volume while in eq.(\ref{fselfd}) a finite 
space-time volume $L^4$ is considered. 

Using (\ref{fselfd}) we numerically evaluate the analytic 
1-loop propagator on a given finite lattice, using
\begin{equation}
{\cal S}_{LL} = {\cal S}^{(0)}_{LL} + {\cal S}^{(0)}_{LL}  
\left[-\Sigma^{I,II}_{LL}\right] {\cal S}^{(0)}_{LL},
\end{equation}
and in the section $IV$ 
compare with nonperturbative numerical results. To avoid the infra-red 
problem in the scalar propagator, we use anti-periodic boundary 
condition in one of the space-time directions in evaluating 
(\ref{fselfd}). 

In a similar
way, 1-loop corrected $RR$ or $LL$ propagators are obtained at all the
$s$-slices of the waveguide boundaries $I$ and $II$, {\em i.e.}, at the
slices $s_0$, $s_0+1$, $s_1$ and $s_1+1$.

\subsection{Fermion mass matrix at tree level and 1-loop} \label{ferm_m}
Another issue of interest is the spread of the
wavefunctions of the two chiral zero mode solutions along the discrete
$s$-direction and their possible overlap. A finite overlap would
mean an induced Dirac mass. The extra dimension, as already pointed out
in the discussion following (\ref{par5}), can be interpreted as a flavor
space with one LH chiral fermion, one RH chiral fermion and $(L_s/2-1)$
heavy fermions on each sector of $1\leq s\leq L_s/2-1$ and $ L_s/2+1\leq
s\leq L_s-1$. 
\subsubsection{Tree level}
For the spread of the zero modes at the tree level, 
one needs only to solve
$M_0 u_L =0$ and $M_0^\dagger u_R=0$ \cite{kaplan,aoki0} where $M_0=
M(p=0)$ (keeping the momentum $p$ non-zero is unnecessary in this
discussion). However, for radiative correction on the domain wall mass
$m(s)$, the heavy mode spreads are also needed. Accordingly we consider
flavor diagonalization of $M^\dagger_0 M_0$ ($M_0$ is not hermitian) as
in \cite{aoki2,blum}:
\begin{eqnarray}
\left( M^\dagger_0 M_0 \right)_{st} \left( \phi^{(0)}_j \right)_t &=& \left[
\lambda_j^{(0)} \right]^2 \left( \phi^{(0)}_j \right)_s \label{eveq1}\\
\left( M_0 M^\dagger_0 \right)_{st} \left( \Phi^{(0)}_j \right)_t &=& \left[
\lambda_j^{(0)} \right]^2 \left( \Phi^{(0)}_j \right)_s. \label{eveq2}
\end{eqnarray}

The index $j$ for the eigenvalues and the eigenvectors is basically a
flavor index, but unlike $s$ and $t$ which vary from $0$ to $L_s-1$, it is
taken symmetric around the domain wall $j=0$. Index $j$ varies from
$-L_s/2$ to $L_s/2$. From periodic boundary condition, $L_s/2$ and 
$-L_s/2$  are the same point in flavor space corresponding to the 
anti-domain wall. It is to be noted that $j$ appears 
explicitly in the heavy mode solutions below. However, it may be pointed 
out that the index $j$ need not be chosen this way. One could also 
define $j$ in the same way as the flavor indices $s$ and $t$, only in 
that case the explicit solutions below would have a different 
appearance and to our taste less tractable. 

To solve the above tree level eigenequations (later also at 1-loop), we 
should ideally use the Kaplan boundary conditions because we used in our
action the Kaplan way of implementing the domain wall. However, that would
make an explicit calculation of the heavy eigenmodes quite complicated and
almost intractable. We have hence devised a special set of boundary
conditions (look at Appendix A for details on the boundary conditions) 
with the property that
no information can be passed through the wall and the antiwall. This 
makes the calculation less cumbersome and explicit expressions can be 
obtained in manageable forms. We stress that these are not 
the actual boundary conditions in the particular domain wall 
implementation we have taken. However, we show in Appendix B that using 
the correct boundary conditions ({\em i.e.}, the Kaplan boundary 
conditions) also would lead to the same qualitative 
conclusions about the nature of 1-loop corrections to the eigenvalues, 
stability of the zero modes, and particularly the 1-loop overlap of the 
opposite chiral zero modes. 
  
With our special boundary conditions (used only in sections \ref{ferm_m}
and \ref{ferm_m1}), the explicit form of the 
eigenfunctions are obviously different from \cite{aoki2,blum} and are 
obtained as, 

\vspace{0.3cm}
\leftline{Zero ($L$-handed) mode:}
\begin{eqnarray}
(j=0) \hspace{2.0cm}
\left( \phi^{(0)}_j \right)_s &=& {\cal A}\;
\exp(-\tilde{\alpha}s) \hspace{1.9cm} 0\leq s\leq L_s/2 \\
(j=L_s/2) \hspace{2.7cm}
&=& {\cal A}\; \exp\left[-\tilde{\alpha}(L_s-s)\right] 
\hspace{0.8cm} L_s/2\leq s\leq L_s \\
{\cal A} &=& \left(
\frac{1-\exp(-2\tilde{\alpha})}{1-\exp(-\tilde{\alpha}L_s)}
\right)^{1/2} \nonumber \\
{\rm with}, \hspace{2.0cm} \exp(-\tilde{\alpha}) &=& 1-m_0
\end{eqnarray}

\leftline{Heavy mode $(\vert j \vert \neq 0,\,L_s/2)$:}
\begin{eqnarray}
(0<j<L_s/2) \hspace{1.0cm} \left( \phi^{(0)}_j \right)_s
&=& \sqrt{\frac{4}{L_s}} \,\sin\beta_j(s-1)
\hspace{2.1cm} 0\leq s\leq L_s/2 \\
(-L_s/2<j<0) \hspace{2.0cm} &=& \sqrt{\frac{4}{L_s}} 
\,\sin\beta_{-j}(L_s/2-s+1) \hspace{0.7cm} L_s/2\leq s\leq L_s .
\end{eqnarray}
The parameters involved are as follows:
\begin{eqnarray}
\left(\lambda^{(0)}_j \right)^2 &= & 1 + \tilde{a}(s)^2 
-2\tilde{a}(s)\cos \beta_j, \label{eigv} \\  
\tilde{a}(s) & = & 1 + m(s) \label{apm} 
\end{eqnarray}
and for nontrivial heavy mode solutions, 
$\beta_j = 2\pi j/L_s$ with $\vert j \vert \ne 0,\,L_s/2$.

Solutions to (\ref{eveq2}) for all $j$ and $s$ easily follow:
\begin{equation}         
\left( \Phi^{(0)}_j \right)_s = \left(
\phi^{(0)}_j \right)_{L_s/2-s}, \label{modes}
\end{equation}

It is obvious that the
eigenvectors $\phi^{(0)}_{j=0}$ and $\Phi^{(0)}_{j=0}$ correspond 
respectively to the
LH and RH chiral zero modes at the domain wall and the anti-domain wall
in the region $0\leq s\leq L_s/2$. Similarly the
eigenvectors $\phi^{(0)}_{j=L_s/2}$ and $\Phi^{(0)}_{j=L_s/2}$ correspond to
the same chiral zero modes at the domain wall and the anti-domain wall 
in the region
$L_s/2\leq s \leq L_s$. The $\vert j \vert \neq 0,\,L_s/2$ eigenvectors 
are for the heavy flavor modes. All solutions are real.

The overlap of the opposite chiral modes in the region $0\leq s\leq L_s/2$
does not depend explicitly on $s$ and is given at the tree level by,
\begin{equation}
\Phi^{(0)}_{j=0}\; \phi^{(0)}_{j=0} = {\cal A}^2\;
\exp(-\tilde{\alpha}L_s/2) 
\end{equation}
and similarly in the region $L_s/2\leq s\leq L_s$. The exponentially 
damped mixing of the LH and RH chiral modes does not induce any Dirac 
mass at the tree level for large $L_s$.
 
It is also noted that $\phi^{(0)}$ and 
$\Phi^{(0)}$ diagonalize $M_0$ and $M_0^\dagger$ in the following 
manner: 
\begin{equation}
\left( \Phi^{(0)}_j\right)_s \left( M_0 \right)_{st}
\left(\phi^{(0)}_{j^\prime} \right)_t = \lambda_j^{(0)} \delta_{j,j^\prime}
= \left( \phi^{(0)}_j\right)_s \left(M_0^\dagger\right)_{st}
\left(\Phi^{(0)}_{j^\prime} \right)_t   \label{diag}
\end{equation}
The way the formalism is set up, the absolute sign of $\lambda_j^{(0)}$ 
in eq.(\ref{diag}) is arbitrary.

\subsubsection{1-loop}
Flavor off-diagonal {\em tadpole} diagrams
produce the self-energies for the $LR$ and $RL$ parts of the fermion
propagator. Again, the
self-energies are nonzero only at the waveguide boundaries $I$ and $II$.

\vspace{-0.7cm}
\begin{center}
\begin{figure}
\begin{picture}(300,85)(0,65)
\ArrowLine(90,85)(150,85)
\ArrowLine(150,85)(210,85)
\DashArrowArcn(150,105)(20,180,0){3}
\DashCArc(150,105)(20,180,360){3}
\Text(95,92)[]{\footnotesize{$L$}}
\Text(112,76)[]{$p$, $s_0+1$}
\Text(205,92)[]{\footnotesize{$R$}}
\Text(198,76)[]{$p$, $s_0$}
\Text(150,115)[]{$k$}
\end{picture}
\caption{Tadpole contribution to 1-loop $LR$ propagator at $WG$ 
boundary $I$.} \label{tadp}
\end{figure}
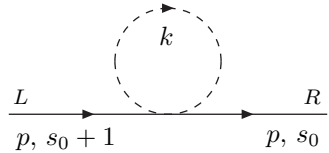
\end{center}

For the $LR$ propagator connecting $s_0$ and $s_0+1$ at the waveguide
boundary $I$, the self-energy contribution from the tadpole diagram is
given by,
\begin{eqnarray}
-\left(\Sigma_{LR}^I(p)\right)_{st} &=& \frac{1}{2}b^2P_L \int_{BZ}
\frac{d^4k}{(2\pi)^4} \frac{1}{\hat{k}^2 (\hat{k}^2+\omega^2)}\;
\delta_{s,s_0}\delta_{t,s_0+1} \\
&=& \frac{1}{2}b^2P_L \,{\cal T}
\;\delta_{s,s_0}\delta_{t,s_0+1}  \label{tadpld}
\end{eqnarray}
where ${\cal T}\sim 0.04$ is the tadpole loop integral.

Similarly the self-energy contribution to the $LR$ propagator at the
waveguide boundary $II$ connecting $s_1$ and $s_1+1$ comes from a
tadpole diagram and is given by,
\begin{equation}
-\left(\Sigma_{LR}^{II}(p)\right)_{st}
= \frac{1}{2}b^2P_L \,{\cal T}
\;\delta_{s,s_1}\delta_{t,s_1+1}.  \label{tadpld2}
\end{equation}

The mass parameter $M_0$ gets modified at 1-loop as:
\begin{eqnarray}
(M_0)_{st}P_L \rightarrow (\widetilde{M}_0)_{st}P_L &=&
(M_0)_{st}P_L+\left[-(\Sigma^I_{LR}(0))_{st} \right] +
\left[-(\Sigma^{II}_{LR}(0))_{st}\right]   \nonumber \\
&\equiv & (M_0)_{st}P_L + b^2 ({\bf \Sigma}_{LR})_{st} P_L\,.
\end{eqnarray}
$\Sigma^{I,\,II}_{LL,RR}(0) =0$ identically. $(M_0^\dagger)_{st}P_R$
gets modified accordingly. 

Because of our use of the special boundary conditions (Appendix A) which do
not allow any communication through the walls, 1-loop correction to 
the eigenvectors and eigenvalues involves either 
$s_0$ (waveguide boundary $I$) or $s_1$ (waveguide boundary $II$) 
depending on which sector of $j$ is considered. In Appendix B we show that 
if the actual (Kaplan) boundary conditions are used, one gets contribution 
at 1-loop level also from another diagram, called there the {\em global
loop} diagram. However, the conclusions are qualitatively the same.  

\subsection{Mass matrix diagonalization at 1-loop} \label{ferm_m1}
At 1-loop level we organize the corrections to the eigenvectors and the
eigenvalues of the fermion mass matrix squared as follows:
\begin{eqnarray}
\phi^{(0)} \rightarrow \phi &=& (1 + b^2 \phi^{(1)})\phi^{(0)}, 
\nonumber \\
\Phi^{(0)} \rightarrow \Phi &=& (1 + b^2 \Phi^{(1)})\Phi^{(0)}, 
\nonumber \\
\left(\lambda^{(0)}\right)^2 \rightarrow \lambda^2 & = &
\left(\lambda^{(0)} + b^2 \lambda^{(1)}\right)^2.
\end{eqnarray}
$\lambda^{(1)}$  is found to be
\begin{equation}
\lambda^{(1)}_j = \left(\Phi^{(0)}_j\right)_s ({\bf
\Sigma}_{LR})_{st} \left(\phi^{(0)}_j\right)_t
\end{equation}
The eq.(\ref{diag}) gets modified in 1-loop as
\begin{equation}
(\Phi_j)_s (\widetilde{M}_0)_{st} (\phi_{j^\prime})_t =
\lambda_j \,\delta_{j,j^\prime} + {\cal O}(b^4) =
(\phi_j)_s (\widetilde{M}_0^\dagger)_{st}
(\Phi_{j^\prime})_t.
\end{equation}

The 1-loop correction to the eigenvalues, 
$(\delta \lambda)_j=b^2 \lambda^{(1)}_j$,  is given by:
\begin{eqnarray}
(\delta \lambda)_j &=& \frac{2b^2}{L_s}\,{\cal T} \sin\beta_j(L_s/2-s_0-1)
\sin\beta_js_0 + {\cal O}(b^4) \hspace{2.0cm} {\rm for}\;\;1\leq j \leq
(L_s/2-1),   \label{delm1}  \\
 &=& \frac{2b^2}{L_s}\,{\cal T} \sin\beta_j(s_1+1) 
\sin\beta_j(L_s/2-s_1) + {\cal O}(b^4) \hspace{1.75cm} {\rm 
for}\;\;-(L_s/2-1) \leq j \leq -1, \label{delm2} \\  
&=& \frac{b^2}{2}\;
{\cal A}^2\,{\cal T} \exp[-\tilde{\alpha} (L_s/2+1)]+ {\cal O}(b^4) 
\;\;\;\;\stackrel{L_s\rightarrow \infty} {\longrightarrow} \;0 
\hspace{1.2cm} {\rm for}\;\;j=0,L_s/2. \label{delm3}
\end{eqnarray}
We notice in 
eq.(\ref{delm3}) that the 1-loop correction to zero mode eigenvalue is 
exponentially damped and the zero modes are hence perturbatively stable.
 
The 1-loop expression for the chiral zero mode at the domain wall, in 
the region $0\leq s\leq L_s/2$, is,
\begin{equation}
(\phi_{j=0})_s = {\cal A} \left[ \exp(-\tilde{\alpha} s) -
\frac{2b^2}{L_s}{\cal T} \exp(-\tilde{\alpha} (s_0+1))
\sum_{j^\prime=1}^{L_s/2-1} \frac{1}{\lambda^{(0)}_{j^\prime}}
\sin\beta_{j^\prime}(L_s/2-s_0-1)\sin\beta_{j^\prime}(s-1) \right].
\end{equation}
We obtain a similar 1-loop expression for the chiral mode at the 
anti-domain wall. 

The overlap of the opposite chiral modes in the region $0\leq 
s\leq L_s/2$ at 1-loop is given by,
\begin{eqnarray}
\Phi_{j=0}\;\phi_{j=0} = {\cal A}^2 \exp(-\tilde{\alpha}L_s/2)
&-& \left(\frac{2b^2}{L_s}\right){\cal A}^2{\cal T}\,{\cal F}_1 \,
\exp[-\tilde{\alpha}(s_0+1)]
\nonumber\\
&-& \left(\frac{2b^2}{L_s}\right){\cal A}^2{\cal T}\,{\cal F}_2 \,
\exp[-\tilde{\alpha}(L_s/2-s_0)], \label{mixing}
\end{eqnarray}
where,
\begin{eqnarray*}
{\cal F}_1 &=& \sum_{j^\prime,s} \frac{1}{\lambda^{(0)}_{j^\prime}}
\exp[-\tilde{\alpha}(L_s/2-s)]
\sin\beta_{j^\prime}(s-1) \sin\beta_{j^\prime}(L_s/2-s_0-1) \\
{\cal F}_2 &=& \sum_{j^\prime,s} \frac{1}{\lambda^{(0)}_{j^\prime}} 
\exp(-\tilde{\alpha}s)
\sin\beta_{j^\prime}s_0 \sin\beta_{j^\prime}(L_s/2-s-1).
\end{eqnarray*}
Eq.(\ref{mixing}) clearly shows that 1-loop corrections to the mixing of 
the LH and RH modes are also exponentially damped. This guards against 
any induced Dirac mass in the domain wall model for large enough $L_s$ 
and the waveguide boundaries $I$ and $II$ chosen approximately 
equidistant from the domain wall and the anti-domain wall.   
In this context we point out that in the Smit-Swift model a shift 
symmetry for the singlet fermion ensures that no fermion mass counter 
term is needed.

Using the actual Kaplan boundary conditions would make the 
relatively nice explicit expressions in eqs.(\ref{delm1} -- \ref{mixing})
much less manageable, to say the least. Diagonalization of the fermion mass
matrix upto 1-loop  using the Kaplan boundary condition is discussed in
Appendix B. 
   
\section{Numerical Results} \label{numr}
\begin{center}
\begin{figure}
\hspace{0.1cm}\psfig{figure=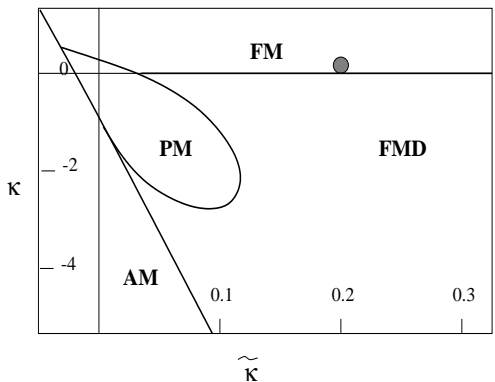,width=6.5cm,height=5.0cm}\\

\caption{Schematic quenched phase diagram.} \label{phases}
\end{figure}
\end{center}

In the quenched approximation, we have first numerically confirmed the
phase diagram in \cite{bock3} of the reduced model in
($\kappa,\tilde{\kappa}$) plane. The phase diagram shown schematically in
Fig.\ref{phases} has the interesting feature that for large enough
$\tilde{\kappa}$, there is a continuous phase transition between the
broken phases FM and FMD. FMD phase is
characterized by loss of rotational invariance and the continuum limit is to
be taken from the FM side of the transition. In the full theory with 
gauge fields, the gauge symmetry reappears at this transition and the gauge
boson mass vanishes, but the longitudinal gauge {\em dof} remain decoupled.
In Fig.\ref{phases} PM is the symmetric phase and AM is the broken
anti-ferromagnetic
phase. The numerical details involved in reconstruction of the phase 
diagram and the fermionic measurements that follow will be 
available in \cite{big}. 

\vspace{0.1cm}

For calculating the fermion propagators, as in \cite{bock1} we have 
chosen the point $\kappa =0.05$, $\tilde{\kappa}=0.2$ (gray blob in 
Fig.\ref{phases}). Although this point is far away from $\tilde{\kappa}=\infty$, 
around which we did our perturbation theory in the previous section, the 
important issue here is to choose a point near the FM-FMD transition and 
away from the FM-PM transition. The results below show that for the 
fermion propagators there is 
excellent agreement between numerical results obtained at $\kappa =0.05$, 
$\tilde{\kappa}=0.2$ and perturbation theory. 
   
Numerically  on $4^3 16$ and $6^3 16$ 
lattices with $L_s=22$ and $m_0=0.5$ we look for chiral modes at the 
domain wall ($s=0$), the anti-domain wall ($s=11$), and at the waveguide 
boundaries ($s=5,6$ and $s=16,17$). Error bars in all the figures are 
smaller than the symbols. 

\begin{figure}
\vspace{-1.0cm}
\parbox{8cm}{\psfig{figure=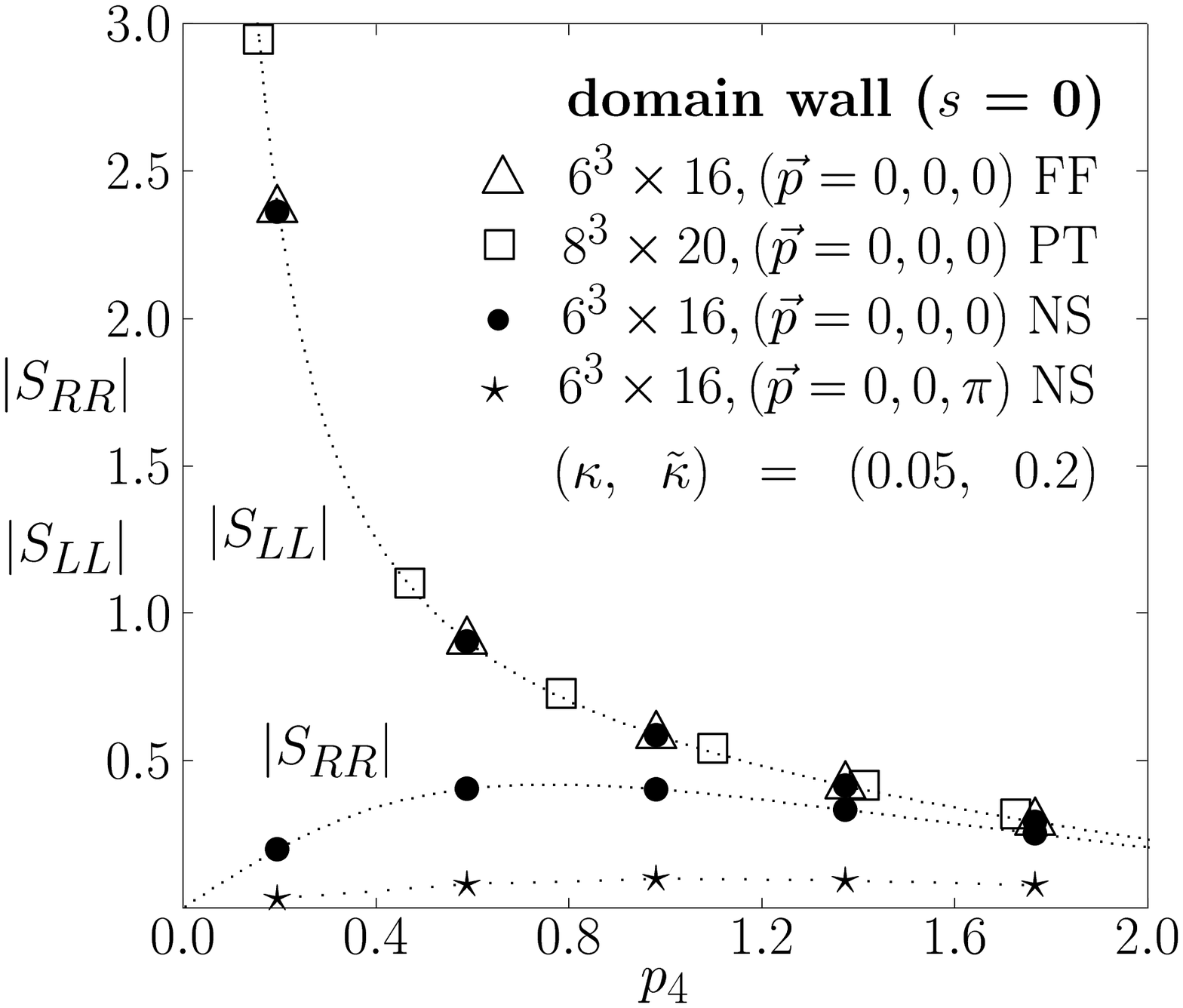,width=8.6cm,height=10.6cm}} \ \
\parbox{8cm}{\psfig{figure=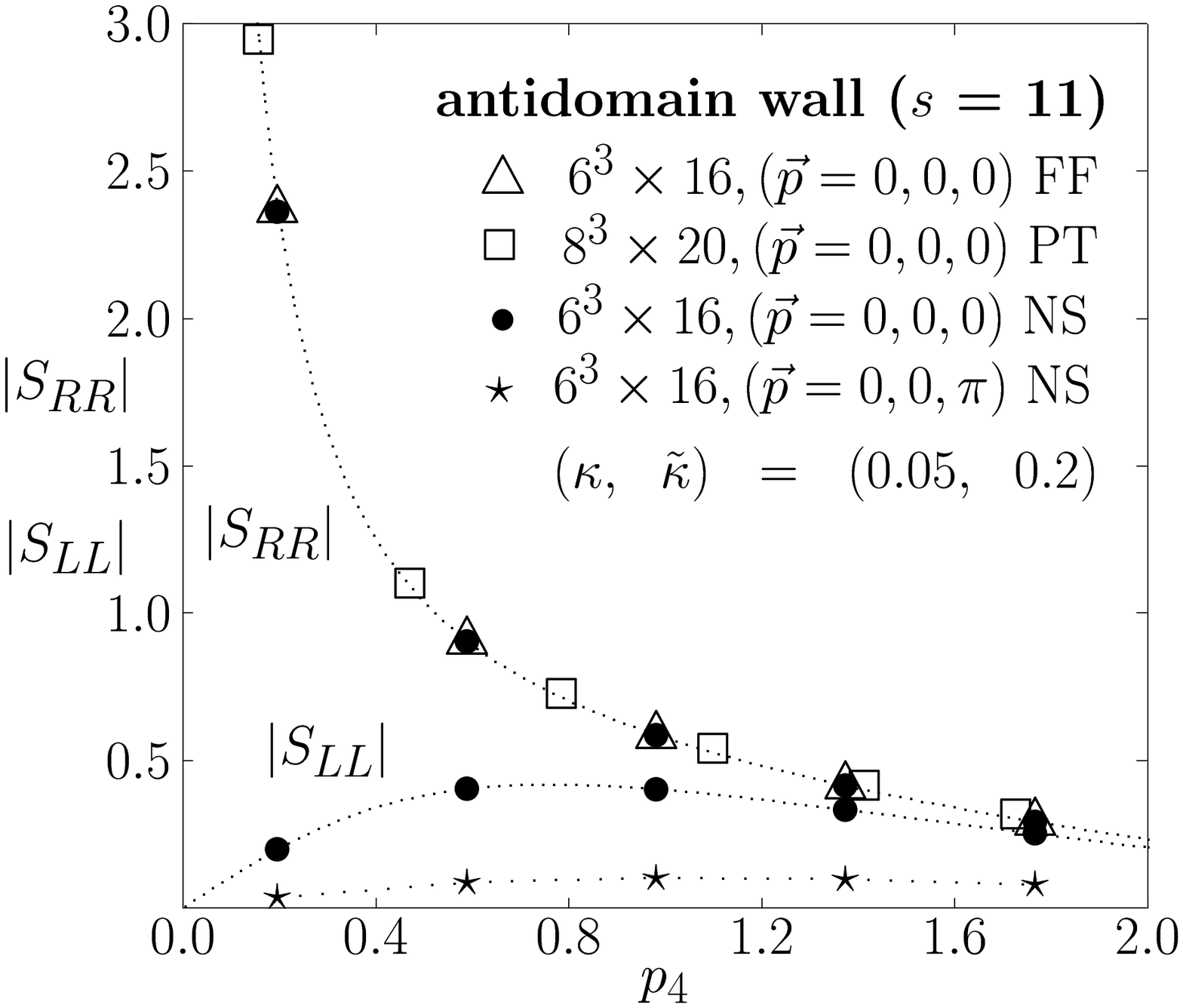,width=8.6cm,height=10.6cm}}
\vspace{-2.7cm}
\caption{Chiral propagators at domain wall $s=0$ and at anti-domain
wall $s=11$ ($L_s=22$; a.p.b.c. in $L_4$, $y=1.0$).} \label{waw}
\end{figure}

Fig.\ref{waw} shows the $RR$ propagator $|S_{RR}|$ and the $LL$ propagator
$|S_{LL}|$ at the domain and anti-domain wall as a function of a component of
momentum $p_4$ for both $\vec{p}=(0,0,0)$ (physical mode) and $(0,0,\pi)$
(first doubler mode) at $y=1$. From the figures, it is clear that
the doubler does not exist, only the physical $RR$ ($LL$) propagator seems to
have a pole at $p=(0,0,0,0)$ at the anti-domain (domain) wall. In all the
figures, NS, PT and FF respectively indicate data from numerical simulation,
from perturbation theory and from free fermion propagator by direct inversion
of the free fermion matrix.

\vspace{0.5cm}

\begin{figure}
\vspace{-1.5cm}
\parbox{8cm}{\psfig{figure=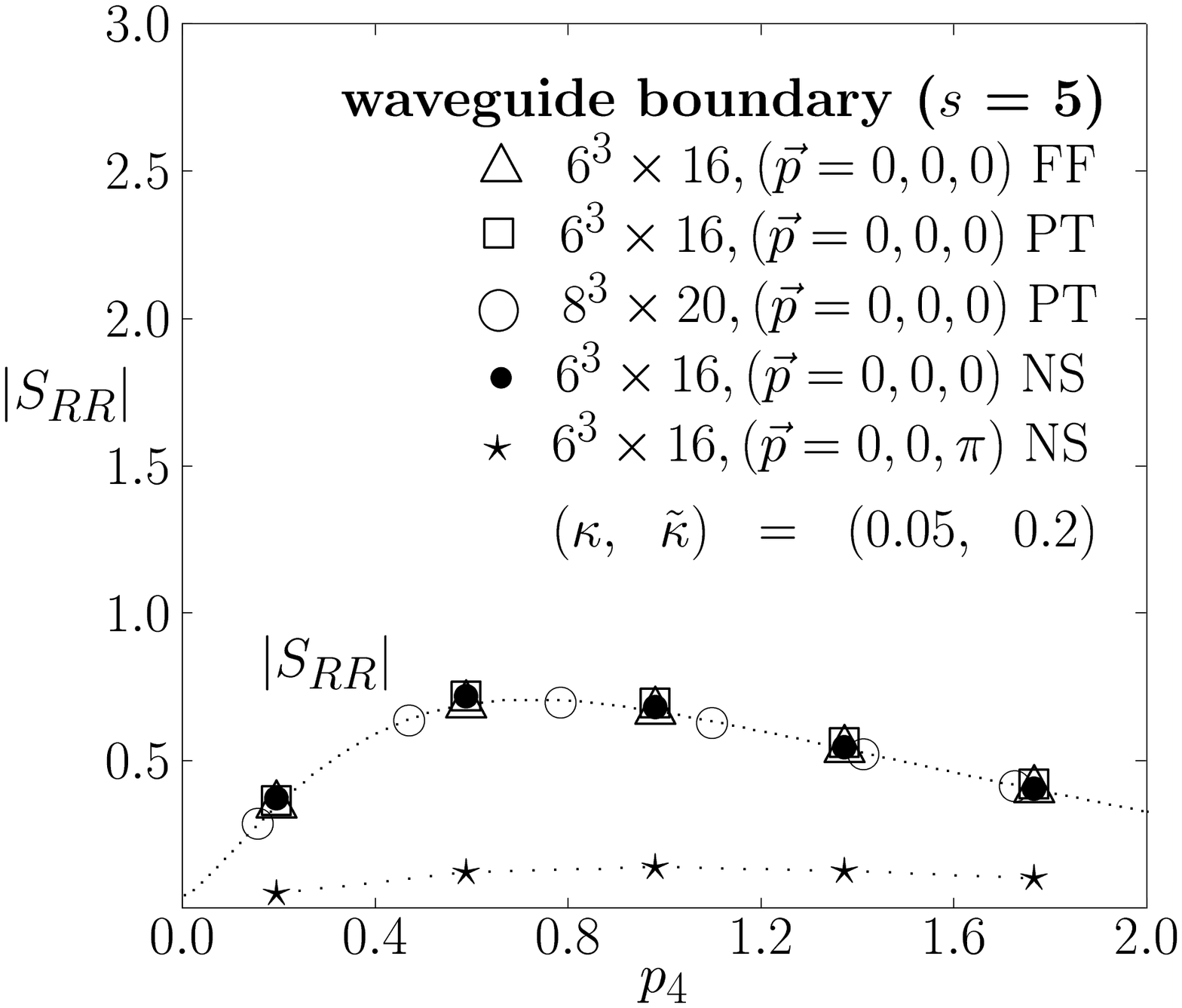,width=8.6cm,height=10.6cm}} \ \
\parbox{8cm}{\psfig{figure=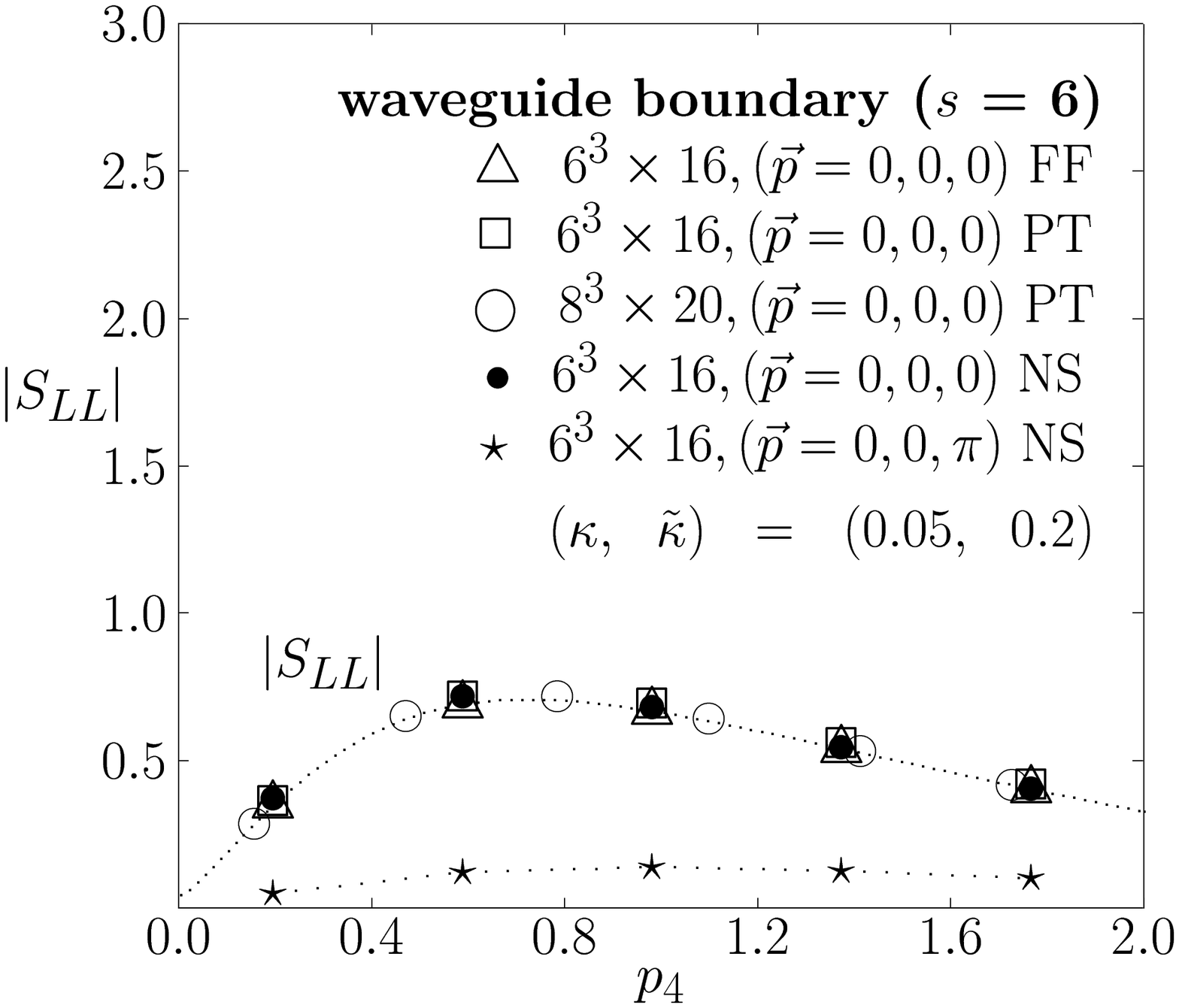,width=8.6cm,height=10.6cm}}
\vspace{-2.9cm}
\caption{$RR$ propagator at waveguide boundary $s=5$ and
$LL$ propagator at waveguide boundary $s=6$ ($L_s=22$; a.p.b.c. in
$L_4$, $y=1.0$).} \label{wgb}
\end{figure}

For Fig.\ref{waw}, PT also means zeroth order perturbation theory, {\em i.e.},
numerical solution of propagator following eq.(\ref{mateq}) (as noted
before in subsubsection \ref{ferm_p1} the self-energy contributions to
the $LL$ and $RR$ propagators are nonzero only at the waveguide boundaries,
the propagators in Fig.\ref{waw} do not get any 1-loop correction).
We have PT
results also for $6^3\times 16$ lattice but have chosen not to show them
because they fall right on top of the numerical data. Instead PT results are
shown for $8^3\times 20$ lattice for which the $p_4$ points are distinct. The
dotted line in all figures refer to the propagator from PT using a
$256^3\times 1024$ lattice. {\em The curves stay the same irrespective of
methods or lattice size}. Based on the above, we can conclude that there are
{\em only free RH fermions} at the anti-domain wall, and at the domain wall
there are {\em only free LH fermions}.

Fig.\ref{wgb} show no evidence of a chiral mode at the waveguide boundaries
$s=5$ and 6 and excellent agreement with 1-loop perturbation
theory. Here too doublers do not exist. Actually the agreement with the FF
method (direct inversion of the free domain wall fermion matrix on a given
finite lattice) is also excellent, because the 1-loop corrections are almost
insignificant. For clarity, in Fig.\ref{wgb} we have not shown the
$LL$ propagator on $s=5$ and the $RR$ propagator at $s=6$, but conclusions are
the same.

Similar investigation at the other waveguide boundary $s=16, 17$ also does
not show any  chiral modes. Previous investigations of the domain wall 
waveguide model without gauge fixing \cite{golter2} have shown that the 
waveguide boundaries are the most likely places to have the unwanted 
mirror modes. This is why we have mostly concentrated in showing that 
there are no mirror chiral modes at these boundaries, although we have 
looked for chiral modes everywhere along the flavor dimension. In fact, 
we do not see any evidence of a chiral mode anywhere other than at the 
domain wall and the anti-domain wall. 

\section{Discussion} \label{disc}

We have followed the gauge fixing proposal of Shamir and Golterman and 
applied it to domain wall fermions for a $U(1)$ $L\chi GT$. By switching 
off the transverse gauge {\em dof}, we arrive at the so-called  
reduced model. We have determined the quenched phase diagram of the 
model and confirmed that there is a continuous phase transition from a 
broken symmetry ferromagnetic (FM) phase to a broken symmetry 
rotationally noninvariant ferromagnetic directional (FMD) phase with the 
properties that at this transition the longitudinal gauge {\em dof}, 
{\em i.e.}, the $\varphi$ fields get decoupled. We have come to this 
conclusion by performing a WCPT for the fermion propagators and 
comparing them to nonperturbative numerical simulations.

Let us now contrast this with the previous attempt \cite{golter2} of 
lattice regularizing a chiral gauge theory with domain wall fermions. No 
gauge fixing was done and in the reduced model, a combination of 
analytic and numerical methods showed that mirror chiral modes were 
dynamically generated in the theory.

As in the Smit-Swift model \cite{smit}, the mirror chiral modes are a 
reflection of the undesired presence of the longitudinal gauge {\em 
dof} in the continuum limit. Without gauge fixing, these radially 
frozen group-valued scalar $\varphi$ fields are generally 
nonperturbative or rough for any value of the transverse gauge coupling, 
even in the reduced model limit. If the continuum limit is taken at the 
FM to a symmetric paramagnetic (PM) phase transition, the $\varphi$ 
fields survive with radial modes and physical effects of them coupling 
with the rest of the theory become manifest in the form of mirror chiral 
modes etc..

In the gauge fixing approach of Shamir and Golterman, care has been 
exercised so that the gauge fixing term is not just a naive lattice 
transcription of the continuum gauge fixing condition. There are 
appropriate additions of irrelevant terms in the covariant gauge fixing 
term so that a unique perturbative vacuum exists. With this gauge 
fixing in the reduced model limit the $\varphi$ fields 
become smooth and can be perturbatively expanded as $1 + {\cal 
O}(1/\sqrt{\tilde{\kappa}}) + \cdots$ around $\tilde{\kappa}=\infty$. 
We have found in our investigation that as long as $\tilde{\kappa}$ is 
taken sufficiently large so that a continuum limit exists from the FM 
phase to the FMD phase (not the PM phase), the model is as good as at 
the perturbative limit $\tilde{\kappa}=\infty$. This seems to hold true 
in our particular implementation, {\em i.e., the domain wall fermion 
case}, in a {\em strong sense}, because the perturbative 1-loop 
corrections ({\em i.e.}, the leading nontrivial corrections) to the 
fermion propagators are found to be negligible (Figs. 4 and 5). In gauge 
fixing the Smit-Swift model, however, 1-loop corrections to the fermion 
propagators were small but not negligible \cite{bock1}. In the present 
investigation with domain wall fermions, to the accuracy of our 
calculations, the reduced model spectrum is that of a free domain wall 
model with absolutely no trace of the $\varphi$ fields.

In our investigation of the reduced model, only a counterterm with 
coefficient $\kappa$ was sufficient to reach the FM-FMD phase transition 
by decreasing $\kappa$ from the FM-side. A dimension-three fermion mass 
counterterm was not needed, because domain wall fermions have the robust 
property that mixing between the opposite chiral modes at the domain 
wall and the anti-domain wall is exponentially damped and is 
negligible for large $L_s$. 

We have carried out and presented the 
perturbative results for fermion propagators and mass matrix in 
reasonable detail, because, although the technique employed is not new, 
the explicit results are available mostly for only the QCD (wall) 
implementation rather than the wall-antiwall implementation suitable for 
a chiral gauge theory. With transverse gauge fields back on (and 
fermions in an anomaly-free representation), the full gauge-fixed domain 
wall model of $L\chi GT$ is ready for a perturbative treatment with the 
same techniques as used in this paper if the transverse gauge coupling 
is perturbative. The model is also ready for a nonperturbative treatment 
by numerical simulation for strong transverse gauge coupling. 

Our numerical computations have been done in the quenched approximation 
and they agree for the fermion propagators very well with zeroth order 
perturbation theory with the 1-loop corrections almost negligible in 
our case. Effects of fermion loops would enter these calculations at 
least at the 2-loop level.  Inclusion of dynamical fermions in our 
nonperturbative numerical investigation should not affect our results 
about the spectrum of the model at all if the relevant part of the phase 
diagram remains qualitatively the same.
 
So far the gauge fixing method is applicable only to the abelian theory. 
Extension to a nonabelian gauge group is nontrivial and  is 
being pursued at the current time \cite{nab}.

As commented at the end of section II, the reduced gauge fixed 
domain wall model  with $0<y<1$ is interesting because at $y=0$ the 
model is known to have mirror modes at the waveguide boundaries. This 
will be taken up in a separate publication \cite{big}.

\acknowledgements 
 
The authors thank Yigal Shamir, Maarten Golterman, Palash Pal and 
Krishnendu Mukherjee for useful discussions. One of the authors (SB) 
would like to thank the Theory Group of Saha Institute of Nuclear 
Physics for providing the facilities.

\appendix
\section{}

In this appendix, we describe the boundary conditions used to 
flavor-diagonalize eqs.(\ref{eveq1}) and (\ref{eveq2}) to obtain the
results presented in subsection \ref{ferm_m} and in \ref{ferm_m1}. 
The boundary conditions make sure that no information passes through the 
domain wall and the anti-domain wall. This means that the eigensolutions 
corresponding to a given source flavor index $s$ belonging to a particular 
segment $0\le s\le L_s/2$ or $L_s/2\le s\le L_s$ will be restricted to that 
particular segment. As already stated before, these are 
not the actual boundary conditions under which the numerical simulations or 
the analytic calculation for the $LL/RR$ chiral propagators have been 
performed, for the flavor diagonalization this enables us to calculate the 
heavy mode solutions and overlap of the chiral modes explicitly. 
In appendix B we have schematically described how the calculations would go 
in case of the actual Kaplan boundary conditions, but to obtain explicit 
expressions would be a horrendous task.  

We consider eq.(\ref{eveq1}) in the region $0\le s,t \le L_s/2$. Obviously
the index $j$ is also in the same region. Writing eq.(\ref{eveq1}) at
$s=L_s/2$ gives (by force using the general expression for $M_0^{\dagger}M_0$
away from the boundaries of the region),
\begin{equation}
\left[1+\tilde{a}(L_s/2)^2 - \left(\lambda^{(0)}_j\right)^2 \right] 
\left(\phi^{(0)}_j\right)_{L_s/2}
- \tilde{a}(L_s/2)\left(\phi^{(0)}_j\right)_{L_s/2+1}
- \tilde{a}(L_s/2-1)\left(\phi^{(0)}_j\right)_{L_s/2-1} =0. \label{AppA1}
\end{equation}
Again writing eq.(\ref{eveq1}) at $s=L_s/2$ gives (this time using the 
expression 
for $M_0^{\dagger}M_0$ at the boundary of the region and dropping terms that 
take the solutions across the anti-domain wall),
\begin{equation}
\left[1+\tilde{a}(L_s/2)^2 - \left(\lambda^{(0)}_j\right)^2 \right] 
\left(\phi^{(0)}_j\right)_{L_s/2}
- \tilde{a}(L_s/2-1)\left(\phi^{(0)}_j\right)_{L_s/2-1} =0. \label{AppA2}
\end{equation}
For both the eqs.(\ref{AppA1}) and (\ref{AppA2}) to be true, the following 
must be true:
\begin{equation}
\left(\phi^{(0)}_j\right)_{L_s/2+1} = 0, \label{AppAbc1}
\end{equation}
which is a boundary condition.

Two equations corresponding to the eqs.(\ref{AppA1}, \ref{AppA2}), this
time at $s=0$, are:
\begin{equation}
\left[1+\tilde{a}(0)^2 - \left(\lambda^{(0)}_j\right)^2 \right] 
\left(\phi^{(0)}_j\right)_{0}
- \tilde{a}(0)\left(\phi^{(0)}_j\right)_{1}
- \tilde{a}(-1)\left(\phi^{(0)}_j\right)_{-1} =0, \label{AppA3}
\end{equation}

\begin{equation}
\left[\tilde{a}(0)^2 - \left(\lambda^{(0)}_j\right)^2 \right] 
\left(\phi^{(0)}_j\right)_{0}
- \tilde{a}(0)\left(\phi^{(0)}_j\right)_{1} =0. \label{AppA4}
\end{equation}

Hence we get the boundary condition at $s=0$ to be
\begin{equation}
\left(\phi^{(0)}_j\right)_0 - \tilde{a}(L_s-1) 
\left(\phi^{(0)}_j\right)_{L_s-1} = 0. \label{AppAbc2}
\end{equation}

For the same eigenequation eq.(\ref{eveq1}) in the other region 
$L_s/2 \le s,t \le L_s$, we similarly get two more boundary conditions as 
follows:
\begin{eqnarray}
\left(\phi^{(0)}_j\right)_{L_s/2} - \tilde{a}(L_s/2-1) 
\left(\phi^{(0)}_j\right)_{L_s/2-1} & = & 0, \label{AppAbc3} \\
\left(\phi^{(0)}_j\right)_1 & = & 0. \label{AppAbc4}
\end{eqnarray}

These boundary conditions are similar to those given by Shamir
\cite{shamir}, except that we get two sets of them, {\em i.e.} 
(\ref{AppAbc1}), (\ref{AppAbc2}), (\ref{AppAbc3}) and (\ref{AppAbc4}),
corresponding to two sectors of $s$. 
Another couple of sets of boundary conditions are obtained for the 
eigenequation eq.(\ref{eveq2}) in the two regions in the same way.

\section{}

Here we schematically show that using the correct boundary conditions 
(in this case
the Kaplan boundary conditions, instead of the special boundary
conditions explained in Appendix A) does 
not qualitatively change our conclusions about fermion mass 
counterterms. 

The use of special boundary  boundary conditions in sec. \ref{ferm_m}
meant that the 5-th or flavor 
dimension was basically divided by the domain and the antidomain 
wall into two segments which did not communicate to each other. As a 
result a wave incident on a wall got fully reflected.
In general, there will be transmission to the other side of 
the wall.  

The chiral zero modes are the same as before (except for a tiny
change in the 
normalization) except that they are now allowed to decay all 
the way around the $s$-space. 

Let us assume then the following trial 
eigenfunctions $\phi^\pm$ and $\Phi^\pm$ for the heavy modes:
\begin{eqnarray}
(\phi^{\pm^{(0)}}_j)_s &=& A_{\pm j} \exp(-i\beta_{\pm j} s) + 
                 B_{\pm j} \exp(i\beta_{\pm j} s) \\
(\Phi^{\pm^{(0)}}_j)_s &=& C_{\pm j} \exp(-i\beta_{\pm j} s) + 
                 D_{\pm j} \exp(i\beta_{\pm j} s).
\end{eqnarray} 
The subscripts $\pm$ go with the regions in $s$-space with 
corresponding signs of $m(s)$ in eq.(\ref{dwmass}). The wavevectors 
$\beta$ are obtained by trying these trial solutions in the eigenvalue 
eqs. (\ref{eveq1}), (\ref{eveq2}) and are given as:
\begin{equation}
2 \tilde{a}(s) \cos\beta_{\pm j} = 1 + \tilde{a}(s)^2 -\lambda_j^2
\end{equation}

We proceed essentially the same way as in the case with the free 
propagators except for the fact that here we do not have the 
inhomogeneous part of the solutions. The boundary conditions used 
to determine $A_{\pm j},\; B_{\pm j},\; C_{\pm j},\; D_{\pm j}$ are
(omitting the subscript $j$ and also the tree level 
indicating superscript $(0)$ for convenience): 
\begin{eqnarray}
\tilde{a}_0 \phi^+_{L_s} - \tilde{a}_- \phi^-_0 &=& 0\\
\tilde{a}_0 \phi^-_{L_s/2} - \tilde{a}_+ \phi^+_{L_s/2} &=& 0\\
(1+\tilde{a}_0^2-\lambda^2) \phi^+_{L_s} - \tilde{a}_0 \phi^-_1 -
\tilde{a}_+ \phi^+_{L_s-1}
&=& 0 \\
(1+\tilde{a}_0^2-\lambda^2) \phi^-_{L_s/2} - \tilde{a}_0 \phi^+_{L_s/2+1} 
- \tilde{a}_- \phi^-_{L_s/2-1} &=& 0 \\
 && \nonumber \\
\tilde{a}_0 \Phi^-_0 - \tilde{a}_+ \Phi^+_{L_s} &=& 0\\
\tilde{a}_0 \Phi^+_{L_s/2} - \tilde{a}_- \Phi^-_{L_s/2} &=& 0\\
(1+\tilde{a}_0^2-\lambda^2) \Phi^-_0 - \tilde{a}_0 \Phi^+_{L_s-1} - 
\tilde{a}_- \Phi^-_1
&=& 0 \\
(1+\tilde{a}_0^2-\lambda^2) \Phi^+_{L_s/2} - \tilde{a}_0 \Phi^-_{L_s/2-1} 
- \tilde{a}_+ \Phi^+_{L_s/2+1} &=& 0              
\end{eqnarray}
where $\tilde{a}_\pm = 1\pm m_0$ and $\tilde{a}_0=\tilde{a}_{L_s/2}=1$ from
eq.(\ref{apm}).

This leads to two sets of linear homogeneous equations for the 
amplitudes:
\begin{eqnarray}
{\bf P}\cdot \Lambda &=& 0 \nonumber \\
{\bf Q} \cdot \Theta &=& 0 \label{pq}
\end{eqnarray}
where ${\bf P}$ and $ {\bf Q}$ are $4\times 4$ matrices and 
$\Lambda = (A_-,\;B_-,\;A_+,\;B_+)$ and $ \Theta 
=(C_-,\;D_-,\;C_+,\;D_+)$. For nontrivial solutions, $\det {\bf P}=0$ 
and $\det {\bf Q}=0$ leading to (reviving the subscript $j$)
\begin{equation}
\beta_{\pm j} = \frac{2\pi}{L_s}j
\end{equation} 

Solutions to the eqs.(\ref{pq}) are of the following general form 
(dropping the subscripts $j$ again): 
\begin{eqnarray}
B_- &=& f_L^{(1)} A_-, \;\;A_+ = f_L^{(2)} A_-, \;\;B_+ = f_L^{(3)} 
A_- \\ 
D_- &=& f_R^{(1)} C_-, \;\;C_+ = f_R^{(2)} C_-, \;\;D_+ = f_R^{(3)} 
C_-
\end{eqnarray}
where $f_{L,R}^{(1,2,3)}$ are complex numbers with finite magnitude 
(order 1). $A_-$ and $C_-$ are determined from normalizations of the 
respective eigenfunctions and would lead to a ${\cal O}(1/L_s)$ factor.

At 1-loop with the Kaplan boundary conditions,  there is also 
a contribution to the fermion self-energy for the 
$LR$ and $RL$ parts coming from a flavor off-diagonal half-circle 
diagram (we shall call it a {\em global-loop} diagram) where the scalar 
field goes around the flavor space connecting fermions at the waveguide 
boundaries $I$ and $II$.  
The global-loop diagram originates from the fact that the $\varphi$ 
field that couples the fermions at the waveguide boundary $I$ is the 
same $\varphi$ field coupling the fermions at the waveguide boundary 
$II$.

\vspace{-0.5cm}
\begin{center}
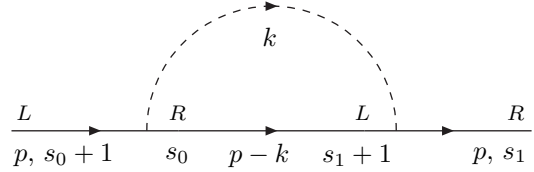
\begin{figure}
\begin{picture}(300,85)(0,65)
\ArrowLine(62,85)(125,85)
\ArrowLine(125,85)(195,85)
\ArrowLine(195,85)(258,85)
\DashArrowArcn(160,85)(47,180,0){3}
\Text(67,92)[]{\footnotesize{$L$}}
\Text(82,76)[]{$p$, $s_0+1$}
\Text(125,92)[]{\footnotesize{$R$}}
\Text(125,76)[]{$s_0$}
\Text(195,92)[]{\footnotesize{$L$}}
\Text(156,76)[]{$p-k$}
\Text(193,76)[]{$s_1+1$}
\Text(253,92)[]{\footnotesize{$R$}}
\Text(247,76)[]{$p$, $s_1$}
\Text(160,121)[]{$k$}
\end{picture}
\caption{Global loop contribution to 1-loop $LR$ propagator connecting $WG$
boundaries $I$ and $II$.} \label{gcirc}
\end{figure}
\end{center}

Self-energy contribution from the global-loop diagram for the $LR$ 
propagator is:  
\begin{eqnarray}
-\left(\Sigma^{gl}_{LR}(p)\right)_{st} &=& b^2 P_L \int_{BZ}
\frac{d^4k}{(2\pi)^4} \left[ M^\dagger(p-k) G_L(p-k) \right]_{s_1+1,s_0}
{\cal G}(k) \;\delta_{s,s_1}\delta_{t,s_0+1} \label{global} \\
&=& b^2 P_L \,{\cal R} \;\delta_{s,s_1}\delta_{t,s_0+1}  \label{nself}
\end{eqnarray}
where ${\cal R}$ is the loop integral in 
eq.(\ref{global}). 

Now the calculations for the 1-loop corrected eigenvalue proceeds 
exactly as in subsection \ref{ferm_m1}, except that it now includes also
the contribution from the global-loop diagram:   
\begin{equation}
\lambda_j = \lambda_j^{(0)} + b^2\,\left(\overline{{\bf \Sigma}}_{LR}
\right)_{jj} 
\end{equation}
where, for $\vert j\vert \ne 0,\,L_s/2$,
\begin{eqnarray}
(\overline{{\bf \Sigma}}_{LR})_{jj}&=&(\Phi^{\pm^{(0)}}_j)_s\,
({\bf \Sigma}_{LR})_{st}\,(\phi^{\pm^{(0)}}_j)_t \\
&=& (\Phi^{\pm^{(0)}}_j)_s\,\left (\frac{1}{2}{\cal T}
(\delta_{s,s_0}\delta_{t,s_0+1}+\delta_{s,s_1}\delta_{t,s_1+1}) 
   +{\cal R}\delta_{s,s_1}\delta_{t,s_0+1}  \right ) 
   (\phi^{\pm^{(0)}}_j)_t \\
&=& \frac{1}{2}{\cal T} \left( 
(\Phi^{-^{(0)}}_j)_{s_0}\,(\phi^{-^{(0)}}_j)_{s_0+1} + 
 (\Phi^{+^{(0)}}_j)_{s_1}\,(\phi^{+^{(0)}}_j)_{s_1+1} \right )
 + {\cal R} (\Phi^{+^{(0)}}_j)_{s_1}\,(\phi^{-^{(0)}}_j)_{s_0+1},
\end{eqnarray} 
and similarly for the chiral zero modes.

Using the expressions for the eigenfunctions at the specific values  
$s_0$, $s_1$, $s_0+1$ and $s_1+1$, we arrive at the 1-loop correction to 
the eigenvalues:
\begin{eqnarray}
(\delta \lambda)_j &\sim & \frac{b^2}{L_s}\left(
{\cal T}{\bf\Omega}_j^{(1)}(s_0,f_L^{(1)}, f_R^{(1)}) +
{\cal T}{\bf\Omega}_j^{(2)}(s_1,f_L^{(2)},f_L^{(3)},f_R^{(2)},f_R^{(3)}) + 
{\cal R}{\bf\Omega}_j^{(3)}(s_0,s_1,f_L^{(1)},f_L^{(2)},f_R^{(3)})
\right),\;\vert j\vert\ne 0,\,\frac{L_s}{2}  \\ 
&\sim & 
b^2 {\cal T} \exp [-\tilde{\alpha}(L_s/2+1)] + b^2 {\cal R} 
\exp[-\tilde{\alpha}(s_0+1)-\tilde{\alpha}(L_s/2-s_1)], 
\;\;\;\;j = 0,\,\frac{L_s}{2} 
\end{eqnarray}
 
The details of the functions ${\bf\Omega}_j^{(n)},\;n=1,2,3$ are not 
illuminating. Correction to the zero mode eigenvalue clearly shows the 
exponential damping and for large $L_s$ it is negligible.

The 1-loop chiral zero mode wavefunction at the domain wall is,
\begin{eqnarray}
(\phi_0)_s &=& (\phi^{(0)}_0)_s + b^2 (\phi^{(1)}_0)_j\,(\phi^{(0)}_j)_s 
\\ 
&\sim &  \exp(-\tilde{\alpha}s) \nonumber\\ 
&&-\frac{b^2}{\lambda^{(0)}_j} \left [
\frac{1}{2}{\cal T}(\phi^{(0)}_0)_{s_0+1}\, (\Phi^{(0)}_j)_{s_0}
(\phi^{(0)}_j)_{s} +
\frac{1}{2}{\cal T}(\phi^{(0)}_0)_{s_1+1}\, (\Phi^{(0)}_j)_{s_1}
(\phi^{(0)}_j)_{s} +
{\cal R}(\phi^{(0)}_0)_{s_0+1}\, (\Phi^{(0)}_j)_{s_1}
(\phi^{(0)}_j)_{s} \right ] \label{zmk1} \\ 
&\sim &  \exp(-\tilde{\alpha}s) \nonumber\\ 
&&-\frac{b^2}{2}{\cal T} \exp[-\tilde{\alpha}(s_0+1)]\, {\bf \xi}_1(s)
- \frac{b^2}{2}{\cal T} \exp[-\tilde{\alpha}(s_1+1)]\, {\bf \xi}_2(s)
- b^2{\cal R} \exp[-\tilde{\alpha}(s_0+1)]\, {\bf \xi}_3(s),
\end{eqnarray}
where the well behaved functions ${\bf \xi}_i,\;i=1,2,3$ contain the 
information of the heavy mode wavefunctions and need not be written down 
explicitly (it is to be noted $j \neq 0$ in eq.(\ref{zmk1})).
The 1-loop correction to the zero-mode wavefunction is also 
explicitly damped and for large $L_s$ negligible. 

The 1-loop correction to zero-mode wavefunction at the anti domain wall 
is similarly found to be damped. As a result the overlap of the opposite 
chiral zero modes at the 1-loop level is also exponentially damped.


\end{document}